\documentclass[12pt]{article}
\def\text#1{\mbox{#1}}
\begin{document}
\begin{center}
{\LARGE \textbf{Elliptic Ruijsenaars-Schneider and Calogero-Moser Models
Represented by Sklyanin Algebra and $sl(n)$ Gaudin Algebra }}

\vskip  3em {\large \lineskip  .75em
Kai CHEN,$^{a}$ Heng FAN,$^{b,a}$ Bo-Yu
HOU,$^{a}$ Kang-jie SHI,$^{a}$ Wen-li YANG$^{a}$ and Rui-hong YUE$^{a}$}\\
{\small \textit{$^{a}$ Institute of Modern Physics, Northwest
University, Xi'an 710069, China}}\\
{\small \textit{$^{b}$
Department Of Physics, Graduate School of Science University of
Tokyo, Tokyo 113-0033, Japan}}
\end{center}

\vskip  3em

\begin{abstract}
The relationship between Elliptic Ruijsenaars-Schneider (RS) and
Calogero-Moser (CM) models with Sklyanin algebra is presented. Lax
pair representations of the Elliptic RS and CM are reviewed. For
$n=2$ case, the eigenvalue and eigenfunction for Lam\'{e} equation
are found by using the result of the Bethe ansatz method.
\end{abstract}

\newpage

\section{Introduction}

A general description of classical completely integrable models of $n$
one-dimensional particles with two-body interactions $V(q_{i}-q_{j})$ was
given in Ref. 1). To each simple Lie algebra and choice of interaction, one
can associate a classically completely integrable system$^{1)-4)}$ such as a
rational, hyperbolic, trigonometric or elliptic CM model.

The Lax pair representation (Lax representation) of a system is a direct
method of showing its integrability and the complete set of integrals of
motion can also be constructed easily. The Lax representation and its
corresponding $r$-matrix for rational, hyperbolic and trigonometric $A_{n-1}$
CM models was constructed by Avan el al.$^{2)}$ The Lax representation for the
elliptic CM models was constructed by Krichever$^{5)}$ and the corresponding
$r$-matrix was given by Sklyanin$^{6)}$ and Braden et al.$^{7)} $ There exists
a specific feature in that the $r$-matrices of the Lax representations for
these models turn out to be dynamical (i.e., they depend on the dynamical
variables) and satisfy dynamical Yang-Baxter equations.$^{8),7),9),6)}$

For the dynamical $r$-matrix, the fundamental Poisson algebra of the Lax
operator, whose structural constants are given by a dynamical $r$-matrix, is
generally no longer closed. The quantization problem and its geometrical
interpretation are also difficult. Considering all of these, a non-dynamical
$r$-matrix is found for these systems.$^{10),11)}$ The trigonometric limit of
these results can be found in Ref. 42). We know the Lax representation for a
completely integrable model is not unique. The different Lax representations
of an integrable system are conjugate to each other (for the field system they
are related by a 'gauge' transformation). The corresponding $r$-matrices are
related by a 'gauge' transformation which is the classical dynamical twisting
relation$^{12)}$ between those $r$-matrices.

The RS model is a relativistic generalization of a CM model. It describes a
completely integrable system of $n$ one-dimensional interacting relativistic
particles. It can be related to the dynamics of solitons in some integrable
relativistic field theory.$^{8),13),9),14)}$ Its discrete-time version has
been connected with the Bethe ansatz equation of the solvable statistical
model.$^{15)}$ Recent developments have shown that it can be obtained by a
Hamiltonian reduction of the cotangent bundle of some Lie group,$^{16)}$ and
can be considered as the gauged WZW theory.$^{17)}$ The Lax representation and
its corresponding $r$-matrix for rational, hyperbolic and trigonometric
$A_{n-1}$ type RS models was constructed by Avan et al.$^{2)}$ The Lax
representation for the elliptic RS models was constructed by
Ruijsenaars,$^{18)}$ and the corresponding $r$-matrix was given by
Nijhoff$^{15)}$ and Suris.$^{19)}$ The main difference between the
$r$-matrices of the relativistic (RS) and non-relativistic (CM) models is that
the latter is given in terms of a linear Poisson-Lie bracket, whereas the
former (RS model) is given in terms of a quadratic Poisson-Lie bracket. In
contrast with the dynamical Yang-Baxter equation of the $r$-matrix for the CM
model.$^{16)}$ the generalized Yang-Baxter relation for the quadratic
Poisson-Lie bracket (RS model) with a dynamical $r$-matrix is still an open
problem.$^{20)}$ Moreover, the Poisson bracket of the Lax operator is no
longer closed, and consequently the quantum version of the classical
$L$-operator, has not been constructed. However, as for the CM model, a
different Lax representation which is conjugated to the original one can be
found. The corresponding $r$-matrix changes by a 'gauge' transformation. The
resulting $r$-matrix may be non-dynamical. Such a transformation may be called
the classical dynamical twisting of the associated linear Poisson-Lie bracket.
Due to the quadratic Poisson-Lie bracket of the RS model, there exist
dynamical twisting relations between the $r$-matrices of Lax operators related
by gauge transformations. Such dynamical twisting is the semi-classical limit
of the quantum dynamical twisting of the $R$-matrix in Ref. 12). For recent
progress in the study of CM models, see, for example, Refs. 21)-24).

The paper is organized as follows: In Sec. 2, we present some
general formulae for dynamical systems. In Sec. 3, we review some
results for the elliptic RS and CM models. The non-dynamical
$r$-matrices for the integrable elliptic systems are then
presented. Their quantization conditions correspond to the quantum
Yang-Baxter relation, and the $R$-matrix is simply the
$Z_{n}$-symmetric Belavin model.$^{28)}$ In Sec. 4, we will
present the relationship between the Sklyanin algebra$^{6),32)}$
and the integrable systems. In Sec. 5, we will obtain the
eigenvalue and eigenfunction for the Lam\'{e} equation. The
Lam\'{e} operator is equivalent to the Hamiltonian of the elliptic
CM model. Section 6 has some brief summary.

\section{The dynamical twisting of classical $r$-matrix}

A Lax pair $(L,M)$ consists of two functions on the phase space of the system
with values in some Lie algebra $g$, such that the evolution equations may be
written in the following form

\begin{eqnarray}
\frac{dL}{dt}=[L,M],
\end{eqnarray}
where [,] denotes the bracket in the Lie algebra $g$. If we have formulated
the Lax pair relation, the conserved quantities (integrals of motion) can be
constructed easily. It follows that the adjoint-invariant quantities
tr$L^{l}(l=1,...,n)$ are the integrals of the motion. In order to implement
the Liouville theorem onto this set of possible action variables we need them
to be Poisson-commuting. As shown in Ref. 25), the commutativity of the
integrals tr$L^{l}$ follows if the Lax operator can be deduced from the
fundamental Poisson bracket

\begin{eqnarray}
\{L_{1}(u),L_{2}(v)\}=[r_{12}(u,v),L_{1}(u)]-[r_{21}(v,u),L_{2}(v)]
\end{eqnarray}
or quadratic form$^{19)}$

\begin{eqnarray}
\{L_{1}(u),L_{2}(v)\}  & =&L_{1}(u)L_{2}(v)r_{12}^{-}(u,v)-r_{21}^{+}
(v,u)L_{1}(u)L_{2}(v)\nonumber\\
&& +L_{1}(u)s^{+}(u,v)L_{2}(v)-L_{2}(v)s^{-}(u,v)L_{1}(u),
\end{eqnarray}
where we use the notation

\begin{eqnarray}
L_{1}=L\otimes1,\mbox{\ \ \ } L_{2}=1\otimes L,\mbox{ \ \ }
a_{21}=Pa_{12}P,
\end{eqnarray}
and $P$ is the permutation operator such that $Px\otimes y=y\otimes x$.

For the above relations to define a consistent Poisson bracket, one should
impose some constraints on the $r$-matrices. The skew-symmetry of the Poisson
bracket requires that

\begin{eqnarray}
r_{21}^{\pm}(v,u)=-r_{12}^{\pm}(u,v),\mbox{ \ \ } s_{21}^{+}(v,u)=s_{12}
^{-}(u,v),
\end{eqnarray}

\begin{eqnarray}
r_{12}^{+}(u,v)-s_{12}^{+}(u,v)=r_{12}^{-}(u,v)-s_{12}^{-}(u,v).
\end{eqnarray}
For the numerical $r$-matrices $r_{12}^{\pm}(u,v)$, $s_{12}^{\pm}(u,v)$, some
constraint conditions (sufficient conditions) are imposed on the $r$-matrix in
order to make it satisfy the Jacobi identity.$^{26)}$ However, generally
speaking, the $r$-matrices $r^{\pm}(u,v)$, $s^{\pm}(u,v)$ depend on dynamical
variables, and the Jacobi identity which implies an algebraic constraint for
the $r$-matrices takes a very complicated form
\begin{eqnarray}
\lbrack L_{1},[r_{12},r_{13}]+[r_{12},r_{23}]+[r_{32},r_{13}]+\{L_{2}
,r_{13}\}-\{L_{3,}r_{12}\}]+\text{cyc.term}=0.
\end{eqnarray}

It should be remarked that for a given integrable system, we can choose
different Lax formulations. The $r$-matrices corresponding to different Lax
formulations are generally different. So, in some cases, we can transform a
dynamical $r$-matrix into a non-dynamical $r$-matrix.$^{10),11)}$ The
different Lax representations of a system are conjugate to each other: if
$(\tilde{L},\tilde{M})$ is another Lax pair of the same dynamical system
conjugate to with the old one $(L,M)$, it means that

\begin{eqnarray}
\frac{d\tilde{L}}{dt}  & =&[\tilde{L},\tilde{M}],\nonumber\\
\tilde{L}{\left(u\right)}  & =&g(u)L(u)g^{-1}(u),\nonumber\\
\tilde{M}(u)  & =&g(u)M(u)g^{-1}(u)-(\frac{d}{dt}g(u))g^{-1}(u),
\end{eqnarray}
where $g(u)\in G$ whose Lie algebra is $g$. Then we have

\textbf{Proposition: }The Lax pair $(\tilde{L},\tilde{M})$. has the following
$r$-matrix structure

\begin{eqnarray}
\{\tilde{L}_{1}(u),\tilde{L}_{2}(v)\}=[\tilde{r}_{12}(u,v),\tilde{L}
_{1}(u)]-[\tilde{r}_{21}(v,u),\tilde{L}_{2}(v)],
\end{eqnarray}
where

\begin{eqnarray}
\tilde{r}_{12}(u,v)  & =&g_{1}(u)g_{2}(v)r_{12}(u,v)g_{1}^{-1}(u)g_{2}
^{-1}(v)+g_{2}(v)\{g_{1}(u),L_{2}(v)g_{1}^{-1}(u)g_{2}^{-1}(v)\nonumber\\
&& +\frac{1}{2}[\{g_{1}(u),g_{2}(v)\}g_{1}^{-1}(u)g_{2}^{-1}(v),g_{2}
(v)L_{2}(v)g_{2}^{-1}(v)].
\end{eqnarray}
\ \ \ \ \ \ \

For a given Lax pair $(L,M)$ and the corresponding $r$-matrix, if there exists
a $g$ such that

\begin{eqnarray}
h_{12}  & =&g_{1}(u)g_{2}(v)r_{12}(u,v)g_{1}^{-1}(u)g_{2}^{-1}(v)+g_{2}
(v)\{g_{1}(u),L_{2}(v)\}g_{1}^{-1}(u)g_{2}^{-1}(v)\nonumber\\
&& +\frac{1}{2}[\{g_{1}(u),g_{2}(v)\}g_{1}^{-1}(u)g_{2}^{-1}(v),g_{2}
(v)L_{2}(v)g_{2}^{-1}(v)]
\end{eqnarray}
and

\begin{eqnarray}
\partial_{q_{i}}h_{12}=\partial_{p_{j}}h_{12}=0
\end{eqnarray}
then a non-dynamical Lax representation of the system exists.

By a straightforward calculation, we can also find that the twisted Lax pair
$(\tilde{L},\tilde{M})$ and the corresponding $r$-matrix $\tilde{r}_{12}$ satisfy

\[
\lbrack\tilde{L}_{1},[\tilde{r}_{12},\tilde{r}_{13}]+[\tilde{r}_{12},\tilde
{r}_{23}]+[\tilde{r}_{32},\tilde{r}_{13}]+\{\tilde{L}_{2},\tilde{r}
_{13}\}-\{\tilde{L}_{3,}\tilde{r}_{12}\}]+\mbox{cycl.term} =0.
\]

Similarly, for the quadratic form, the Lax pair $(\tilde{L},\tilde{M})$ has
the following $r$-matrix structure

\begin{eqnarray}
\{\tilde{L}_{1}(u),\tilde{L}_{2}(v)\}  & =&\tilde{L}_{1}(u)\tilde{L}
_{2}(v)\tilde{r}_{12}^{-}(u,v)-\tilde{r}_{12}^{+}(u,v)\tilde{L}_{1}
(u)\tilde{L}_{2}(v)\nonumber\\
&& +\tilde{L}_{1}(u)\tilde{s}_{12}^{+}(u,v)\tilde{L}_{2}(v)-\tilde{L}
_{2}(v)\tilde{s}_{12}^{-}(u,v)\tilde{L}_{1}(u),
\end{eqnarray}
where

\begin{eqnarray}
\tilde{r}_{12}^{-}(u,v)  & =&g_{1}(u)g_{2}(v)r_{12}^{-}(u,v)g_{1}^{-1}
(u)g_{2}^{-1}(v)-\tilde{\bigtriangleup}_{12}(u,v)+\tilde{\bigtriangleup}
_{21}(v,u),\nonumber\\
\tilde{r}_{12}^{+}(u,v)  & =&g_{1}(u)g_{2}(v)r_{12}^{+}(u,v)g_{1}^{-1}
(u)g_{2}^{-1}(v)-\tilde{\bigtriangleup}_{12}^{(1)}(u,v)+\tilde{\bigtriangleup
}_{21}^{(1)}(v,u),\nonumber\\
\tilde{s}_{12}^{+}(u,v)  & =&g_{1}(u)g_{2}(v)s_{12}^{+}(u,v)g_{1}^{-1}
(u)g_{2}^{-1}(v)-\tilde{\bigtriangleup}_{21}(u,v)-\tilde{\bigtriangleup}
_{12}^{(1)}(u,v),\nonumber\\
\tilde{s}_{12}^{-}(u,v)  & =&g_{1}(u)g_{2}(v)s_{12}^{-}(u,v)g_{1}^{-1}
(u)g_{2}^{-1}(v)-\tilde{\bigtriangleup}_{12}(u,v)-\tilde{\bigtriangleup}
_{21}^{(1)}(v,u),\nonumber\\
\tilde{\bigtriangleup}_{12}(u,v)  &
=&\tilde{L}_{2}^{-1}(v)\bigtriangleup _{12}(u,v),\text{ \
}\tilde{\bigtriangleup}_{12}^{(1)}(u,v)=\bigtriangleup
_{12}(u,v)\tilde{L}_{2}^{-1}(v),\nonumber\\
\bigtriangleup_{12}(u,v)  & =&g_{2}(v)\{g_{1}(u),L_{2}(v)\}g_{1}^{-1}
(u)g_{2}^{-1}(v)\nonumber\\
&& +\frac{1}{2}[\{g_{1}(u),g_{2}(v)\}g_{1}^{-1}(u)g_{2}^{-1}(v),g_{2}
(v)L_{2}(v)g_{2}^{-1}(v)].
\end{eqnarray}
There are still relations:

\begin{eqnarray}
\tilde{r}_{21}^{\pm}(v,u)  &
=&-\tilde{r}_{12}^{\pm}(u,v),~~~~~~~~\tilde
{s}_{21}^{+}(v,u)=\tilde{s}_{12}^{-}(u,v),\nonumber\\
\tilde{r}_{12}^{+}(u,v)-\tilde{s}_{12}^{+}(u,v)  & =&\tilde{r}_{12}
^{-}(u,v)-\tilde{s}_{12}^{-}(u,v).
\end{eqnarray}

And also we have, for given Lax pair $(L,M)$ and the corresponding
$r$-matrices, if there exists a $g$ such that:

\begin{eqnarray}
g_{1}(u)g_{2}(v)s_{12}^{+}(u,v)g_{1}^{-1}(u)g_{2}^{-1}(v)-\tilde{\Delta}
_{21}(v,u)-\tilde{\Delta}_{12}^{(1)}(u,v)   =0,\nonumber\\
g_{1}(u)g_{2}(v)s_{12}^{-}(u,v)g_{1}^{-1}(u)g_{2}^{-1}(v)-\tilde{\Delta}
_{12}(u,v)-\tilde{\Delta}_{21}^{(1)}(v,u)   =0,\nonumber\\
h_{12}(u,v)   =g_{1}(u)g_{2}(v)r_{12}^{-}(u,v)g_{1}^{-1}(u)g_{2}
^{-1}(v)-\tilde{\Delta}_{12}(u,v)+\tilde{\Delta}_{21}^{(1)}(v,u)\nonumber\\
=g_{1}(u)g_{2}(v)r_{12}^{+}(u,v)g_{1}^{-1}(u)g_{2}^{-1}(v)-\tilde{\Delta
}_{12}^{(1)}(u,v)+\tilde{\Delta}_{21}^{(1)}(v,u)
\end{eqnarray}
and

\begin{eqnarray}
\partial_{q_{i}}h_{12}=\partial_{p_{j}}h_{12}=0,
\end{eqnarray}
then a non-dynamical Lax representation with Sklyanin Poisson-Lie bracket for
the system exists.

\section{Lax pair for elliptic RS and CM models}

We first define some elliptic functions:

\begin{eqnarray}
\theta^{(j)}(u) &  =\theta\left[
\begin{array}
[c]{c}
\frac{1}{2}-\frac{j}{n}\\
\frac{1}{2}
\end{array}
\right]  (u,n\tau),~~~\sigma(u)=\theta\left[
\begin{array}
[c]{c}
{\frac{1}{2}}\\
{\frac{1}{2}}
\end{array}
\right]  (u,\tau),\nonumber\\
\theta\left[
\begin{array}
[c]{c}
a\\
b
\end{array}
\right]  (u,\tau) &  =\sum_{n=-\infty}^{\infty}\exp\Big( i\pi\lbrack
(m+a)^{2}\tau+2(m+a)(z+b)]\Big) ,\nonumber\\
\theta^{\prime}{}^{(j)}(u) &  =\partial_{u}\Big( \theta^{(j)}(u)\Big)
,\sigma^{\prime}(u)=\partial_{u}\Big( \sigma(u)\Big) ,\xi(u)=\partial
_{u}\Big( \ln\sigma(u)\Big) ,
\end{eqnarray}
where $\tau$ is a complex number with Im$(\tau)>0$.

The Ruijsenaars-Schneider model is a system of $n$ one-dimensional
relativistical particles interacting by a two-body potential. In terms of the
canonical variables $p_{i},q_{i}$ $(i=1,\cdots,n)$ enjoying the canonical
Poisson bracket

\begin{eqnarray}
\{p_{i},p_{j}\}=0,~~~~\{q_{i},q_{j}\}=0,~~~~\{q_{i},p_{j}\}=\delta_{ij},
\end{eqnarray}
the Hamiltonian of the system is expressed as$^{18)}$

\begin{eqnarray}
H=mc^{2}\sum_{j=1}^{n}\cosh\left(  p_{j}\prod_{k\not =j}\left\{  \frac
{\sigma(q_{jk}+\gamma)\sigma(q_{jk}-\gamma)}{\sigma^{2}(q_{jk})}\right\}
^{\frac{1}{2}}\right)  ,
\end{eqnarray}
where $q_{jk}=q_{j}-q_{k}$, $m$ denotes the particle mass, $c$ the speed of
light, and $\gamma$ is the coupling constant. The above defined Hamiltonian is
known to be completely integrable.$^{18),27)}$ As we mentioned above, the Lax
representation (Lax operator of the classical $L$-operator) is one of the most
effective ways to show that the system is integrable. One Lax representation
for the elliptic RS model was first formulated by Ruijsenaars:$^{18)}$

\begin{eqnarray}
L_{R}(u)_{j}^{i}=\frac{e^{p_{j}}\sigma(q_{ji}+u+\gamma)}{\sigma(\gamma
+q_{ji})\sigma(u)}\prod_{k\not =j}^{n}\left\{  \frac{\sigma(q_{jk}
+\gamma)\sigma(q_{jk}-\gamma)}{\sigma^{2}(q_{jk})}\right\}  ^{\frac{1}{2}
},~~~~i,j=1,\cdots,n.
\end{eqnarray}
Here, we use another Lax representation $\tilde{L}_{R}^{\ 20)}$

\begin{eqnarray}
\tilde{L}_{R}(u)_{j}^{i}=\frac{e^{p_{j}}\sigma(u+q_{ji}+\gamma)}
{\sigma(u)\sigma(q_{ji}+\gamma)}\prod_{k\not =j}\frac{\sigma(\gamma+q_{jk}
)}{\sigma(q_{jk})}.
\end{eqnarray}
$\tilde{L}_{R}^{\ }$ can be obtained from the standard Ruijsenaars' $L_{R}(u)
$ by using a Poisson map

\begin{eqnarray}
q_{i}\longrightarrow q_{i},~~~p_{i}\longrightarrow p_{i}+{\frac{1}{2}
}\ln\prod_{k\not =i}\frac{\sigma(q_{ik}+\gamma)}{\sigma(q_{ik}-\gamma)}.
\end{eqnarray}

Following the work of Nijhoff et al.,$^{20)}$ the fundamental Poisson bracket
of tb Lax operator $\tilde{L}_{R}(u)$ can be given in the following quadratic
$r$-matrix form with dynamical $r$-matrices

\begin{eqnarray}
\{\tilde{L}_{R}(u)_{1},\tilde{L}_{R}(v)_{2}\}  & =&\tilde{L}_{R}(u)_{1}
\tilde{L}_{R}(v)_{2}r_{12}^{-}(u,v)-r_{12}^{+}(u,v)\tilde{L}_{R}(u)_{1}
\tilde{L}_{R}(v)_{2}\nonumber\\
&& +\tilde{L}_{R}(u)_{1}s_{12}^{+}(u,v)\tilde{L}_{R}(v)_{2}-\tilde{L}
_{R}(v)_{2}s_{12}^{-}(u,v)\tilde{L}_{R}(u)_{1},
\end{eqnarray}
where

\[
r_{12}^{-}(u,v)=a_{12}(u,v)-s_{12}(u)+s_{21}(v),\mbox{ \ \ \ } r_{12}
^{+}(u,v)=a_{12}(u,v)+u_{12}^{+}+u_{12}^{-},
\]

\begin{eqnarray}
s_{12}^{+}(u,v)=s_{12}(u)+u_{12}^{+},\mbox{ \ \ \ \ \ \ \ } s_{12}
^{-}(u,v)=s_{21}(v)-u_{12}^{-},
\end{eqnarray}
and

\begin{eqnarray}
u_{12}^{\pm}  & =&\sum_{ij}\xi(q_{ji}\pm \gamma) e_{ii} \otimes e_{jj},\nonumber\\
a_{12}(u,v)  & =&r_{12}^{0}(u,v)+\sum_{i=1}\xi(u-v)e_{ii}\otimes e_{ii}
+\sum_{i\neq j}\xi(q_{ij})e_{ii}\otimes e_{^{jj}},\\
r_{12}^{0}(u,v)  & =&\sum_{i\neq
j}\frac{\sigma(q_{ij}+u-v)}{\sigma
(q_{ij})\sigma(u-v)}e_{ij}\otimes e_{ji},\mbox{ \ \ }
s_{12}(u)=\sum
_{ij}(\tilde{L}_{R}(u)\partial_{\gamma}\tilde{L}_{R}(u)_{j}^{i}e_{ij}\otimes
e_{ji}.
\end{eqnarray}
The following properties are satisfied:

\begin{eqnarray}
&  r_{21}^{\pm}(v,u)=-r_{12}^{\pm}(u,v),\mbox{ \ } s_{21}^{+}(v,u)=s_{12}
^{-}(u,v),\nonumber\\
&  r_{12}^{+}(u,v)-s_{12}^{+}(u,v)=r_{12}^{-}(u,v)-s_{12}^{-}(u,v).
\end{eqnarray}

Here we would like to reformulate the Lax formulation for the RS model. Define
an $n$ $\otimes$ $n$ matrix $A(u;q)$ as:

\begin{eqnarray}
A(u;q)_{j}^{i}\equiv A(u,q_{1},\cdots,q_{n})_{j}^{i}=\theta^{(i)}
(u+nq_{j}-\sum_{k=1}^{n}q_{k}+\frac{n{-1}}{2}).
\end{eqnarray}
Here we should point out that $A(u;q)_{j}^{i}$ corresponds to the intertwiner
function of $\phi_{j}^{(i)}$ between the $Z_{n}$-symmetric Belavin
$R$-matrix$^{28),29)}$ and the $A_{n-1}^{(1)}$ face model.$^{30),31)}$

Define

\begin{eqnarray}
g(u) &  =&A(u;q)\Lambda(q),~~~\Lambda(q)_{j}^{i}=h_{i}(q)\delta_{j}
^{i},\nonumber\\
h_{j}(q) &&  \equiv h_{j}(q_{1},\cdots,q_{N})=\frac{1}{\prod_{l\not =i}
\sigma(q_{il})}.
\end{eqnarray}
We can construct the new Lax operator $L(u)$ as

\begin{eqnarray}
L(u)=g(u)\tilde{L}_{R}(u)g^{-1}(u).
\end{eqnarray}
More explicitly, it can be written as:

\begin{eqnarray}
L(u)_{j}^{i}=\sum_{k=1}^{n}\frac{1}{\sigma(\gamma)}A(u+n\gamma;q)_{k}
^{i}A^{-1}(u;q)_{j}^{k}e^{p_{k}},i,j=1,2,\cdots,n.
\end{eqnarray}
It can be proved that the fundamental Poisson bracket of $L(u)$ can be given
in the quadratic Poisson-Lie form with a nondynamical $r$-matrix:

\begin{eqnarray}
\{L_{1}(u),L_{2}(v)\}=[r_{12}(u-v),L_{1}(u)L_{2}(v)].
\end{eqnarray}
Here the numerical $r$-matrix is the classical $Z_{n}$-symmetric
$r$-matrix.$^{32)}$It takes the form

\begin{eqnarray}
r_{ij}^{lk}(v)=\left\{
\begin{array}
[c]{ll}
(1-\delta_{i}^{l})\frac{\theta^{\prime}{}^{(0)}(0)\theta^{(i-j)}(v)}
{\theta^{(l-j)}(v)\theta^{(i-l)}(0)}+\delta_{i}^{l}\delta_{j}^{k}\left(
\frac{\theta^{\prime}{}^{(i-j)}(v)}{\theta^{(i-j)}(v)}-\frac{\sigma^{\prime
}(v)}{\sigma(v)}\right)   & \mbox{ } {if}~i+j=l+k~\text{mod}
~n \\
0 & {otherwise}
\end{array}
\right.  .
\end{eqnarray}

We know the $Z_{n}$ symmetric $r$-matrix satisfies the nondynamical classical
Yang-Baxter equation

\begin{eqnarray}
\lbrack r_{12}(v_{1}-v_{2}),r_{13}(v_{1}-v_{3})]+[r_{12}(v_{1}-v_{2}
),r_{23}(v_{2}-v_{3})]  \nonumber\\
+[r_{13}(v_{1}-v_{3}),r_{23}(v_{2}-v_{3})]=0,
\end{eqnarray}
We also know that this $r$-matrix is antisymmetric and $Z_{n}$ symmetric:

\begin{eqnarray}
 \mbox{Antisymmetry} :&&-r_{21}(-v)=r_{12}(v),\nonumber\\
  Z_{n}\otimes Z_{n}\mbox{Symmetry} :&&r_{12}(v)=(a\otimes a)r_{12}
(v)(a\otimes a)^{-1},
\end{eqnarray}
where $a =g$, $h$, and $g$, $h$ are $n\otimes n$ matrices defined as:

\begin{eqnarray}
h_{ij}=\delta_{{i+1},j\mbox{ mod }n},~~~g_{ij}=\omega^{i}\delta_{i,j}.
\end{eqnarray}
For convenience, we also define another $n\otimes n$ matrix

\begin{eqnarray}
I_{\alpha}\equiv I_{\alpha_{1},\alpha_{2}}\equiv g^{\alpha_{2}}h^{\alpha_{1}},
\end{eqnarray}
where $\alpha_{1},\alpha_{2}\in Z_{n}$ and $\omega=exp(2\pi i/n)$.

Next, we will consider the non-relativistic limit of the Lax operator $L(u)$.
First rescale the momenta $\{p_{i}\}$, the coupling constant $\gamma$ and the
Lax operator as follows:$^{20)}$

\begin{eqnarray}
p_{i}:=-\beta p_{i}^{\prime},\ \ n\gamma:=\beta s,\ \ L(u):=\sigma
(\frac{\beta s}{n})L^{\prime}(u).
\end{eqnarray}
Here notation $L^{\prime}$ is introduced. The non-relativistic limit is
then obtained by taking the limit $\beta$ $\rightarrow0$. We have the
following asymptotic properties

\begin{eqnarray}
L^{\prime}(u)_{j}^{i}=\delta_{j}^{i}-\beta(\sum_{k}\{A(u;q)_{k}
^{i}A(u;q)_{j}^{k}p_{k}^{\prime}-s\partial_{s(}A(u;q)_{k}^{i}
)A^{-1}(u;q)_{j}^{k}\})+O(\beta^{2}).
\end{eqnarray}
If we make the canonical transformation

\begin{eqnarray}
p_{i}^{\prime}\rightarrow
p_{i}^{\prime}-\frac{s}{n}\frac{\partial }{\partial q_{i}}\ln
M(q),\mbox{ \ \ } M(q)=\prod_{i<j}\sigma (q_{ij}),
\end{eqnarray}
we finally obtain the Lax operator of the elliptic $A_{n-1}$ CM model$^{10)}$

\begin{eqnarray}
L_{CM}(u)_{j}^{i}=-\lim_{\beta\rightarrow0}\frac
{L^{\prime}(u)_{j}^{i}-\delta_{j}^{i}}{\beta}|_{p_{i}^{\prime
}\rightarrow p_{i}^{\prime}-\frac{s}{n}\frac{\partial}{\partial q_{i}
}\ln M(q)}
\end{eqnarray}
Here we have

\begin{eqnarray}
\{L_{CM}(u)_{1},L_{CM}(v)_{2}\}=[r_{12}(u-v),L_{CM}(u)_{1}+L_{CM}(v)_{2}].
\end{eqnarray}

For the newly constructed Lax representation $L(u)$, the quantization becomes
no longer difficult. Define the $Z_{n}$-symmetric Belavin's $R$-matrix as:

\begin{eqnarray}
R_{ij}^{lk}(u)=\left\{
\begin{array}
[c]{ll}
\frac{\theta^{\prime}{}^{(0)}(0)\sigma(u)\sigma(\sqrt{-1}\hbar)}
{\sigma^{\prime}(0)\theta^{(0)}(v)\sigma(v+\sqrt{-1}\hbar)}\frac{\theta
^{(0)}(v)\theta^{(i-j)}(v+\sqrt{-1}\hbar)}{\theta^{(i-l)}(\sqrt{-1}
\hbar)\theta^{(l-j)}(v)}, & \mbox{if} ~i+j=l+k~\mbox{mod}~n,\\
0 & \mbox{otherwise.}
\end{array}
\right.
\end{eqnarray}
We know this $R$-matrix satisfies the quantum Yang-Baxter equation

\begin{eqnarray}
R_{12}(u_{1}-u_{2})R_{13}(u_{1}-u_{3})R_{23}(u_{2}-u_{3})=R_{23}(u_{2}
-u_{3})R_{13}(u_{1}-u_{3})R_{12}(u_{1}-u_{2}).
\end{eqnarray}
The $R$-matrix is $Z_{n}$ symmetric in the sense that

\begin{eqnarray}
R_{12}(u)=(a\otimes a)R_{12}(u)(a\otimes a)^{-1},~~a=g,h.
\end{eqnarray}
By taking the special limit $\hbar\rightarrow0$, we can obtain the classical
$Z_{n}$ symmetric $r$-matrix

\begin{eqnarray}
R_{12}(u)|_{\hbar\rightarrow0}=1\otimes1+\sqrt{-1}\hbar r_{12}(u)+o(\hbar
^{2}).
\end{eqnarray}
Now let us study the quantum $L$-operator, using the usual canonical
quantization procedure

\begin{eqnarray}
p_{j}\rightarrow\hat{p}_{j}=-\sqrt{-1}\hbar\frac{\partial}{\partial q_{j}
},~~~q_{j}\rightarrow q_{j},~~~j=1,\cdots,n.
\end{eqnarray}
The corresponding quantum $L$-operator can be formulated as:

\begin{eqnarray}
\hat{L}(u)_{l}^{m} &
=&\frac{1}{\sigma(\gamma)}\sum_{k=1}^{n}A(u+n\gamma
;q)_{k}^{m}A^{-1}(u;q)_{l}^{k}e^{\hat{p}_{k}}\nonumber\\
&  =&\frac{1}{\sigma(\gamma)}\sum_{k=1}^{n}A(u+n\gamma;q)_{k}^{m}
A^{-1}(u;q)_{l}^{k}e^{-\sqrt{-1}\hbar\frac{\partial}{\partial q_{k}}
}.\label{anquanL}
\end{eqnarray}

It should be remarked that this quantum $L$-operator is simply the factorised
difference representation for the elliptic $L$-operator.$^{31),33)}$ The above
defined quantum $L$-operator satisfies the quantum Yang-Baxter relation

\begin{eqnarray}
R_{12}(u-v)\hat{L}_{1}(u)\hat{L}_{2}(v)=\hat{L}_{2}(v)\hat{L}_{1}
(u)R_{12}(u-v).
\end{eqnarray}
The proof can be found in Refs. 31), 34), 33) and 35).

\section{RS and CM models related with Sklyanin algebra}

We introduce here some notation for elliptic functions:

\begin{eqnarray}
\sigma_{\alpha}(u) &  =&\theta\left[
\begin{array}
[c]{c}
{\frac{1}{2}}+{\frac{\alpha_{1}}{n}}\\
{\frac{1}{2}}+{\frac{\alpha_{2}}{n}}
\end{array}
\right]  (u,\tau),\nonumber\\
W_{\alpha}(u) &
=&\frac{\sigma_{\alpha}(u+\sqrt{-1}\hbar)}{\sigma_{\alpha
}(\sqrt{-1}\hbar)}\frac{\sigma_{0}(\sqrt{-1}\hbar)}{\sigma_{0}(u+\sqrt
{-1}\hbar)}.
\end{eqnarray}
The above mentioned quantum $R$-matrix can be rewritten as following up to a scale:

\begin{eqnarray}
R(u)=\sum_{\alpha}W_{\alpha}(u)I_{\alpha}\otimes I_{\alpha}^{-1},
\end{eqnarray}
as before $\alpha\equiv\alpha_{1},\alpha_{2}$ and$\ \alpha_{i}\in Z_{n},i=1,2$.

The quantum $L$-operator $\hat{L}$ obtained in the last section can be
represented by the generators of Sklyanin algebra S$_{\alpha}$:
\begin{eqnarray}
\hat{L}(u)=\sum_{\alpha}V_{\alpha}(u)I_{\alpha}S_{\alpha},
\end{eqnarray}
where
\begin{eqnarray}
V_{\alpha}(u)=\frac{\sigma_{\alpha}(u^{\prime}+\sqrt{-1}\hbar\xi)}
{n\sigma_{0}(u^{\prime})\sigma_{\alpha}(\sqrt{-1}\hbar)},\mbox{
\ } u^{\prime}=u+n\sqrt{-1}\hbar\delta-\frac{n-1}{2},
\end{eqnarray}
where $\delta$ is a constant.

The quantum Yang--Baxter relation (49) gives the defining relations of the
Sklyanin algebra:$^{32),6)}$

\begin{eqnarray}
\sum_{\gamma}\frac{\omega^{\gamma_{1}-\alpha_{1}+(\beta_{1}-\gamma_{1}
)(\gamma_{1}-\alpha_{2})}\sigma_{\alpha+\beta-2\gamma}(0)\sigma_{\beta}
(2\sqrt{-1}\hbar)}{\sigma_{\gamma-\alpha}(\sqrt{-1}\hbar)\sigma_{\alpha
+\beta-\gamma}(\sqrt{-1}\hbar)\sigma_{\gamma}(\sqrt{-1}\hbar)\sigma
_{\beta-\gamma}(\sqrt{-1}\hbar)}S_{\alpha+\beta-\gamma}S_{\gamma}=0,
\end{eqnarray}
with $\alpha_{i},\beta_{i},\gamma_{i}\in Z_{n},i=1,2$.

We can give a realization of the generators of Sklyanin algebra as:

\begin{eqnarray}
S_{\alpha}=\sum_{j}S_{j\alpha}e^{-n\sqrt{-1}\frac{\partial}{\partial q_{j}}}.
\end{eqnarray}
Here we introduce the symbol $S_{j\alpha}.$ for the elliptic function
\begin{eqnarray}
S_{j\alpha}=(-1)^{\alpha_{1}}\sigma_{\alpha}(\sqrt{-1}\hbar)\prod_{k\not =
j}\frac{\sigma_{\alpha}(\sqrt{-1}\hbar\xi+q_{jk})}{\sigma_{0}(q_{jk})}.
\end{eqnarray}

Next. we will consider the classical limit of the above defining relations.
Letting $\hbar\rightarrow0$, the quantum $R$-matrix become the classical $r
$-matrix, and we explicitly have the elements of the $r$-matrix in (34)
presented in the last section, here we use another notation

\begin{eqnarray}
R(u)=1+\sqrt{-1}\hbar r(u)+O(\hbar^{2}).
\end{eqnarray}
The classical $r$-matrix is written as:

\begin{eqnarray}
r(u)=\sum_{\alpha}w_{\alpha}(u)I_{\alpha}\otimes I_{\alpha}^{-1},
\end{eqnarray}
where

\begin{eqnarray}
w_{0}(u) &  =&0,\nonumber\\
w_{\alpha}(u) &  =&\frac{\sigma_{\alpha}(u)\sigma_{0}^{\prime}(0)}
{\sigma_{\alpha}(0)\sigma_{0}(u)},~~~\alpha\neq0.
\end{eqnarray}
In order to consider the classical limit of $L$, we first present the
classical limit of $V_{\alpha}(u)$:

\begin{eqnarray}
V_{0}(u) &  =&\frac{\sigma_{0}(u^{\prime})+\sqrt{-1}\hbar\xi\sigma_{0}^{\prime
}(u^{\prime})+O(\hbar^{2})}{n\sigma_{0}(u^{\prime})\sigma_{0}(\sqrt
{-1}\hbar)},\\
V_{\alpha}(u) &  =&\frac{\sigma_{\alpha}(u^{\prime})}{n\sigma_{\alpha
}(0)\sigma_{0}(u^{\prime})}+\frac{\sqrt{-1}\hbar}{n\sigma_{0}(u^{^{\prime}
})}\left[  \frac{\xi\sigma_{\alpha}^{\prime}(u^{\prime})\sigma_{\alpha
}(0)-\sigma_{\alpha}(u^{\prime})\sigma_{\alpha}^{\prime}(0)}{\sigma
_{\alpha}^{2}(0)}\right]  +O(\hbar^{2}),\nonumber\\
&&  \alpha\neq0.
\end{eqnarray}
From the definition of the operator $S_{\alpha}$, we easily have

\begin{eqnarray}
S_{\alpha}=\sum_{j}S_{j\alpha}\left(  1-n\sqrt{-1}\hbar\frac{\partial
}{\partial q_{j}}+O(\hbar^{2})\right)  .
\end{eqnarray}
In the limit $\hbar\rightarrow0$, the elliptic functions $S_{j\alpha}$ take
the forms:

\begin{eqnarray}
S_{j0} &  =&\sigma(\sqrt{-1}\hbar)\left[  1+\sqrt{-1}\hbar\xi\sum_{k\not =
j}\frac{\sigma_{0}^{\prime}(q_{jk})}{\sigma_{0}(q_{jk})}+O(\hbar^{2})\right]
,\\
S_{j\alpha} &  =&(-1)^{\alpha_{1}}\sigma_{\alpha}(0)\prod_{k\neq j}\frac
{\sigma_{\alpha}(q_{jk})}{\sigma_{0}(q_{jk})}\left[  1+\sqrt{-1}\hbar\left(
\frac{\sigma_{\alpha}^{\prime}(0)}{\sigma_{\alpha}(0)}+\xi\sum_{k\not =j}
\frac{\sigma_{\alpha}^{\prime}(q_{jk})}{\sigma_{\alpha}(q_{jk})}\right)
+O(\hbar^{2})\right]  ,\nonumber\\
&&  \alpha\neq0.
\end{eqnarray}
So, we have

\begin{eqnarray}
V_{0}(u)S_{0} &  =&1+\sqrt{-1}\hbar\left[  \xi\frac{\sigma_{0}^{\prime
}(u^{\prime})}{\sigma_{0}(u^{\prime})}\right]  +\frac{{\sqrt{-1}\hbar}
}{n}\sum_{j}\left[  \xi\sum_{k\not =j}\frac{\sigma_{0}^{\prime}(q_{jk}
)}{\sigma_{0}(q_{jk})}-n\frac{\partial}{\partial q_{j}}\right]  +O(\hbar
^{2}),\nonumber\\
&  =&1+\sqrt{-1}\hbar\left[  \xi\frac{\sigma_{0}^{\prime}(u^{\prime}
)}{\sigma_{0}(u^{\prime})}\right]  +\frac{{\sqrt{-1}\hbar}}{n}\sum
_{j}\left[  -n\frac{\partial}{\partial q_{j}}\right]  +O(\hbar^{2}
),\label{v0s0}\\
V_{\alpha}(u)S_{\alpha} &  =&\sqrt{-1}\hbar(-1)^{\alpha_{1}}\frac
{\sigma_{\alpha}(u^{\prime})}{n\sigma_{0}(u^{\prime})}\sum_{j}
\prod_{k\not =j}\frac{\sigma_{\alpha}(q_{jk})}{\sigma_{0}(q_{jk})}\left[
\xi\sum_{k\not =j}\frac{\sigma_{\alpha}^{\prime}(q_{jk})}{\sigma_{\alpha
}(q_{jk})}-n\frac{\partial}{\partial q_{j}}\right]  +O(\hbar^{2}),\nonumber\\
&&  +O(\hbar^{2}),~~~~~~~~~~~~~~~~~~~~~~\alpha\neq0.\label{v1s1}
\end{eqnarray}
here we have $\sum_{j}\sum_{k\not =j}\frac{\sigma_{0}^{\prime}(q_{jk})}
{\sigma_{0}(q_{jk})}=0$, $\ $ because $\frac{\sigma_{0}^{\prime}(q_{jk}
)}{\sigma_{0}(q_{jk})}$ is an odd function.

We can finally expand the quantum $\hat{L}$ operator in the order of $\hbar$
when we take a limit $\hbar\rightarrow0$. However, we first introduce some notation

\begin{eqnarray}
\hat{L}(u)=\sum_{\alpha}V_{\alpha}(u)S_{\alpha}I_{\alpha}=1+\sqrt{-1}\hbar
l(u)+O(\hbar^{2}),
\end{eqnarray}
where $l(u)$ is the classical $l$ operator. We may represent $l(u)$ in terms
of generators of the ``classical'' Sklyanin algebra $S_{\alpha}$:

\begin{eqnarray}
l(u)=\frac{\xi\sigma_{0}^{\prime}(u^{\prime})}{n\sigma_{0}(u^{^{\prime}
})}-\sum_{\alpha}v_{\alpha}(u)S_{\alpha}.
\end{eqnarray}
The function $v_{\alpha}(u)$ is defined as:

\begin{eqnarray}
v_{0}(u) &  =&\frac{1}{n},\\
v_{\alpha}(u) &  =&\frac{1}{n}\frac{\sigma_{\alpha}(u^{\prime}
)\sigma_{\alpha}(0)}{\sigma_{0}(u^{\prime})},~~\alpha\neq0.
\end{eqnarray}

From the above obtained results, we can realize the generators of the
``classical" Sklyanin algebra in the following forms:

\begin{eqnarray}
s_{0} &  =&\sum_{j}n\frac{\partial}{\partial q_{j}},\\
s_{\alpha} &  =&(-1)^{\alpha_{1}}\sigma_{\alpha}(0)\sum_{j}\prod_{k\not =
j}\frac{\sigma_{\alpha}(q_{jk})}{\sigma_{0}(q_{jk})}\left[  n\frac{\partial
}{\partial q_{j}}-\xi\sum_{k\not =j}\frac{\sigma_{\alpha}^{\prime}(q_{jk}
)}{\sigma_{\alpha}(q_{jk})}\right]  ,\mbox{ } \alpha\neq0.
\end{eqnarray}
On the other hand, here we can say we give a definition of the generators of
the ``classical'' Sklyanin algebra.

In the classical limit, the quantum Yang-Baxter relation becomes the following

\begin{eqnarray}
\lbrack l_{1}(u),l_{2}(v)]=[r_{12}(u-v),l_{1}(u)+,l_{2}(v)]
\end{eqnarray}
Substitute the $l$-operator with the generators of the ''classical'' Sklyanin
algebra, and through tedious calculation, we have

\begin{eqnarray}
\lbrack s_{\alpha},s_{\gamma}]=(\omega^{\alpha_{1}\gamma_{2}}-\omega
^{\alpha_{2}\gamma_{1}})\left(  \frac{\sigma_{0}^{\prime}(0)}{n}\right)
s_{\alpha+\gamma}.
\end{eqnarray}
On the other hand, we find that $I_{\alpha}$ satisfy a similar relation

\begin{eqnarray}
\lbrack I_{\alpha},I_{\gamma}]=(\omega^{\alpha_{1}\gamma_{2}}-\omega
^{\alpha_{2}\gamma_{1}})I_{\alpha+\gamma}.
\end{eqnarray}
So, after rescaling $s_{\alpha}$ , we find \{$s_{\alpha}\}${.} and
$\{I_{\alpha}\}$ satisfy the same algebra.

Finally we should point out that if we substitute $\frac{\partial}{\partial
q_{k}}$ by the corresponding canonical variable $p_{k}$, the $l$-operator will
become a $T$-operator, and the commutative bracket on the left-hand side of
the above relation (74) will change to the standard Poisson-Lie bracket. Here
we rewrite as:

\begin{eqnarray}
\{T_{1}(u),T_{2}(v)\}=[r_{12}(u-v),T_{1}(u)+T_{2}(v)].
\end{eqnarray}

\section{CM model, Gaudin model, Lam\'{e} equation and the Bethe ansatz}

For the difference factorized operator $\hat{L}$, we can find some commuting
families which are related to conserved operators. By using the fusion
procedure, the commuting family take the form

\begin{eqnarray}
D_{m}=tr[\hat{L}_{(u)}\otimes\cdots\otimes\hat{L}_{(u)}P_{-}^{m}], \nonumber
\end{eqnarray}
there are $m$ $\hat{L}^{'}s\ $above, $P_{-}^{m}$ is the anti-symmetric
projector. In the classical limit, we also have a similar commuting family

\begin{eqnarray}
a_{m}(u)=\sum_{\alpha_{i}\neq0}v_{\alpha_{1}}(u)\cdots v_{\alpha_{m}
}(u)s_{a_{1}}\cdots s_{a_{m}}\text{tr}[I_{\alpha_{1}}\otimes I_{\alpha_{m}
}P_{-}^{m}],
\end{eqnarray}
where $\alpha^{i}\in Z_{n}^{2},i=1,\cdots,m.$ Let $u\prime=0$, so
$u=u_{0}=\frac{n-1}{2}-n\hbar\xi$, and after rescaling $a_{l}(u)$, we have

\begin{eqnarray}
a_{m}(u_{0})=\sum_{\alpha_{i}\neq0}s_{\alpha_{1}}\cdots s_{\alpha_{m}
}\text{tr}[I_{\alpha_{1}}\otimes I_{\alpha_{m}}P_{-}^{m}],
\end{eqnarray}
We will discuss a special case $n=2$, explicitly we have

\begin{eqnarray}
s_{01}  & =&2\sigma_{01}(0) \left(\xi\frac{\sigma_{01}^{\prime}(q_{12})}
{\sigma_{0}(q_{12})}-\frac{\sigma_{01}(q_{12})}{\sigma_{0}(q_{12})}
(\frac{\partial}{\partial q_{1}}-\frac{\partial}{\partial q_{2}}
)\right),\nonumber\\
s_{10}  & =&2\sigma_{10}(0)\left(\xi\frac{\sigma_{10}^{\prime}(q_{12})}
{\sigma_{0}(q_{12})}-\frac{\sigma_{10}(q_{12})}{\sigma_{0}(q_{12})}
(\frac{\partial}{\partial q_{1}}-\frac{\partial}{\partial q_{2}}
)\right),\nonumber\\
s_{11}  & =&2\sigma_{11}(0)\left(\xi\frac{\sigma_{11}^{\prime}(q_{12})}
{\sigma_{0}(q_{12})}-\frac{\sigma_{11}(q_{12})}{\sigma_{0}(q_{12})}
(\frac{\partial}{\partial q_{1}}-\frac{\partial}{\partial q_{2}})\right).
\end{eqnarray}
We will calculate the non-trivial conserved operator

\begin{eqnarray}
4\alpha_{2}(u)=\sum_{a\neq0}\frac{\sigma_{\alpha}(u^{\prime})\sigma_{-\alpha
}(u^{\prime})}{\sigma_{\alpha}(0)\sigma_{-\alpha}(0)}s_{\alpha}s_{-\alpha
}\omega^{-\alpha_{1}\alpha_{2}}.
\end{eqnarray}
After some tedious calculations, we have

\begin{eqnarray}
\alpha_{2}  & =&-\xi^{2}\frac{\sigma_{0}^{^{\prime\prime}}(q)}{\sigma_{0}
(q)}+2\xi\frac{\sigma_{0}^{\prime}(q)}{\sigma_{0}(q)}(\frac{\partial
}{\partial q_{1}}-\frac{\partial}{\partial q_{2}})+4\frac{\partial^{2}
}{\partial q_{1}\partial q_{2}}\nonumber\\
&& -(\frac{\partial}{\partial q_{1}}+\frac{\partial}{\partial q_{2}})^{2}
-(\xi^{2}+2\xi)[\frac{\sigma_{0}^{\prime}(u^{\prime})^{2}}{\sigma
_{0}(u^{\prime})^{2}}-\frac{\sigma_{0}^{^{\prime\prime}}(u^{\prime}
)}{\sigma_{0}(u^{\prime})}],
\end{eqnarray}
where $q=q_{1}-q_{2}$. This relation is just the same as that obtained by
Hasegawa.$^{36)}$

Since $1$ and $\frac{\partial}{\partial q_{1}}+\frac{\partial}{\partial q_{2}
}$ are also conserved quantities defined above, after some tedious
calculations we have another conserved operator

\begin{eqnarray}
H=\frac{\partial^{2}}{\partial q^{2}}-\xi\frac{\sigma_{0}^{\prime}
(q)}{\sigma_{0}(q)}\frac{\partial}{\partial q}+\frac{\xi^{2}\sigma
_{0}^{^{\prime\prime}}(q)}{4\sigma_{0}(q)}.
\end{eqnarray}
We can change it to a more familiar form. Let $\xi=2\beta$, and suppose $\psi$
and $\Lambda$ are an eigenfunction and eigenvalue of the above Hamiltonian

\begin{eqnarray}
H{\psi}=\Lambda\psi.
\end{eqnarray}
At the same time, we introduce a transformation of this eigenfunction
$\psi=\tilde{\psi}[\sigma_{0}(q)]^{\beta}$, we thus have the following relations:

\begin{eqnarray}
H\tilde{\psi}[\sigma_{0}(q)]^{\beta}=[\frac{d^{2}}{dq^{2}}-2\beta\frac
{\sigma_{0}^{\prime}(q)}{\sigma_{0}(q)}\frac{d}{dq}+\beta^{2}\frac
{\sigma_{0}^{^{\prime\prime}}(q)}{\sigma_{0}(q)}]\tilde{\psi}[\sigma
_{0}^{\beta}(q)]^{\beta}=\Lambda\tilde{\psi}[\sigma_{0}^{\beta}(q)]^{\beta}.
\end{eqnarray}
This means:

\begin{eqnarray}
H^{\prime}\tilde{\psi}=[\frac{d^{2}}{dq^{2}}+\beta(\beta+1)\frac{d^{2}
}{dq^{2}}\ln\sigma_{0}(q)]\tilde{\psi}=\Lambda\tilde{\psi}.
\end{eqnarray}
One finds that $H^{\prime}$ is simply the Hamiltonian of the CM model, see,
for example, Refs. 3), 4), 37) and 38). It is also connected to the Lam\'{e}
operator, see Ref. 39) and the references therein.

Next, we will calculate the eigenfunction and eigenvalue of the above defined
Lam\'{e} operator $H$. Here we first review some of the results obtained by
Felder and Varchenko.$^{39)}$ The difference operator $L$ which is equivalent
to $S_{0}$, one generator of the Sklyanin algebra when $n=2$, is given by

\begin{eqnarray}
L\psi(q)=\frac{\sigma_{0}(q+2\hbar\beta)}{\sigma_{0}(q)}\psi(q-2\hbar
)+\frac{\sigma_{0}(q-2\hbar\beta)}{\sigma_{0}(q)}\psi(q+2\hbar).
\end{eqnarray}
This difference operator is also called the $q$-deformed Lam\'{e} operator. In
the frame work of the quantum inverse scattering method, there is a result as
follows:$^{39)}$

Let $(t_{1},\cdots,t_{m},c)$ be a solution of the Bethe ansatz equations:

\begin{eqnarray}
\frac{\sigma_{0}(t_{i}-2\hbar\beta)}{\sigma_{0}(t_{i}+2\hbar\beta)}
\prod_{j\neq i}\frac{\sigma_{0}(t_{j}-t_{i}-2\hbar)}{\sigma_{0}(t_{j}
-t_{i}+2\hbar)}=e^{4\hbar c},\text{ \ }i=1,\cdots,\beta,
\end{eqnarray}
such that $t_{i}\neq t_{j}$ mod $Z+\tau Z$ if $i\neq j$. Then

\begin{eqnarray}
\psi(q)=e^{cq}\prod_{j}^{\beta}\sigma_{0}(q+t_{j}),
\end{eqnarray}
is a solution of the $q$-deformed Lam\'{e} equation $L\psi=\epsilon\psi$ with eigenvalue

\begin{eqnarray}
\epsilon=e^{-2\hbar c}\frac{\sigma_{0}(4\hbar\beta)}{\sigma_{0}(2\hbar\beta
)}\prod_{j=1}^{\beta}\frac{\sigma_{0}(t_{j}+(2\beta-2)\hbar)}{\sigma_{0}
(t_{j}+2\beta\hbar)}.
\end{eqnarray}
By taking the special limit $\hbar\rightarrow0$, the difference operator becomes:

\begin{eqnarray}
L=2+4\hbar^{2}\left[\frac{d^{2}}{dq^{2}}-2\beta\frac{\sigma_{0}^{\prime}
(q)}{\sigma_{0}(q)}\frac{d}{dq}+\beta^{2}\frac{\sigma_{0}^{^{\prime\prime}
}(q)}{\sigma_{0}(q)}\right]+O(\hbar^{4}).
\end{eqnarray}
We find the term with order $\hbar^{2}$ is exactly the Hamiltonian presented
in relation (85)

\begin{eqnarray}
H=\frac{d^{2}}{dq^{2}}-2\beta\frac{\sigma_{0}^{\prime}(q)}{\sigma_{0}
(q)}\frac{d}{dq}+\beta^{2}\frac{\sigma_{0}^{^{\prime\prime}}(q)}{\sigma
_{0}(q)}.
\end{eqnarray}
Since we already know the eigenfunction of the difference operator $L$ is
$\psi(q)=e^{cq}\prod_{j=1}^{\beta}\sigma_{0}(q+t_{j})$ which does not depend
on $\hbar$, we need only expand the eigenvalue of $L$ in the order of $\hbar$.
We can obtain the eigenvalue of the Lam\'{e} operator $H$

\begin{eqnarray}
\epsilon & =&2-4\hbar\left[  c+\sum_{j=1}^{\beta}\frac{\sigma_{0}^{^{\prime}
}(t_{j})}{\sigma_{0}(t_{j})}\right]  +8\hbar^{2}c\left[  c+\sum_{j=1}^{\beta
}\frac{\sigma_{0}^{\prime}(t_{j})}{\sigma_{0}(t_{j})}\right]  \nonumber\\
&& -4\hbar^{2}c^{2}+4\beta^{2}\hbar^{2}\frac{\sigma_{0}^{^{^{\prime\prime
\prime}}}(0)}{\sigma_{0}^{\prime}(0)}+4\hbar^{2}\left\{  \sum_{j=1}^{\beta
}\left[\frac{\sigma_{0}^{\prime}(t_{j})}{\sigma_{0}(t_{j})}\right]^{2}+2\sum
_{i>j}\frac{\sigma_{0}^{\prime}(t_{i})\sigma_{0}^{\prime}(t_{j})}
{\sigma_{0}(t_{i})\sigma_{0}(t_{j})}\right\}  \nonumber\\
&& +4\hbar^{2}(1-2\beta)\sum_{j=1}^{\beta}\left[  \frac{\sigma_{0}
^{^{\prime\prime}}(t_{j})}{\sigma_{0}(t_{j})}-\left(\frac{\sigma_{0}^{^{\prime}
}(t_{j})}{\sigma_{0}(t_{j})}\right)^{2}\right]  +O(\hbar^{4}).
\end{eqnarray}
At the same time we take the limit $\hbar\rightarrow0$ for the Bethe ansatz
equation, obtaining

\begin{eqnarray}
c+\beta\frac{\sigma_{0}^{\prime}(t_{i})}{\sigma_{0}(t_{i})}-\sum_{j,j\neq
i}\frac{\sigma_{0}^{\prime}(t_{i}-t_{j})}{\sigma_{0}(t_{i}-t_{j})}=0
\end{eqnarray}
Considering this Bethe ansatz equation, we can finally find the eigenvalue of
the Lam\'{e} operator $\Lambda$,

\begin{eqnarray}
\Lambda=(1-2\beta)\left[  \sum_{j=1}^{\beta}\frac{\sigma_{0}^{\prime}
(t_{j})}{\sigma_{0}(t_{j})}\right]  ^{\prime}+\beta^{2}\frac{\sigma
_{0}^{^{^{\prime\prime\prime}}}(0)}{\sigma_{0}^{\prime}(0)}.
\end{eqnarray}
Here we have the results:

Let $(t_{1},\cdots,t_{m},c)$ be a solution of the Bethe ansatz equations:

\begin{eqnarray}
c+\beta\frac{\sigma_{0}^{\prime}(t_{i})}{\sigma_{0}(t_{i})}-\sum_{j,j\neq
i}\frac{\sigma_{0}^{\prime}(t_{i}-t_{j})}{\sigma_{0}(t_{i}-t_{j})}=0,\text{
\ \ }i=1,\cdots,\beta,
\end{eqnarray}
such that $t_{i}\neq t_{j}$ mod $Z+\tau Z$ if $i\neq j$. Then

\begin{eqnarray}
\psi(q)=e^{cq}\prod_{j}^{\beta}\sigma_{0}(q+t_{j})
\end{eqnarray}
is a solution of the equation

\begin{eqnarray}
H\psi(q)=\left[\frac{d^{2}}{dq^{2}}-2\beta\frac{\sigma_{0}^{\prime}(q)}
{\sigma_{0}(q)}\frac{d}{dq}+\beta^{2}\frac{\sigma_{0}^{^{\prime\prime}}
(q)}{\sigma_{0}(q)}\right]\psi(q)=\Lambda\psi(q),
\end{eqnarray}
with eigenvalue

\begin{eqnarray}
A=(1-2\beta)\left[\sum_{j=1}^{\beta}\frac{\sigma_{0}^{\prime}(t_{j})}{\sigma
_{0}(t_{j})}\right]^{\prime}+\beta^{2}\frac{\sigma_{0}^{^{^{\prime\prime\prime}}
}(0)}{\sigma_{0}^{\prime}(0)}.
\end{eqnarray}
We can also obtain these results by directly using the algebraic Bethe ansatz
method for the $A_{1}^{(1)}$ Gaudin model,$^{40)}$ just like the algebraic
Bethe ansatz for the XYZ Gaudin model.$^{41)}$

\section{Summary}
We review some developments concerning the non-dynamical structure of the
elliptic RS and CM models. We also give a solution to the Lam\'{e} equation.
The eigenfunction and eigenvalue for the Lam\'{e} operator are found through
the results of the Bethe ansatz.

The results of the last sections are only for the $n=2$ case. For general $n$,
we can also obtain the eigenvalue and eigenfunction for the generalized Lam\'{e}
operator by using the algebraic Bethe ansatz method for the $sl(n)$ elliptic
Gaudin model. The conserved quantities also correspond to the Hamiltonian of
the elliptic CM model.

\section*{Acknowledgements}

H. Fan is supported by the Japan Society for the Promotion of Science. He
would like to thank the hospitality of Professor Wadati's group in the
Department of Physics, University of Tokyo. B. Y. Hou would like to thank the
hospitality of Professor Sasaki and the Yukawa Institute in Kyoto University.
This work is supported in part by the NSFC, Climbing project, NWU teachers'
fund and Shaanxi province's fund.


\vspace{1pt}

\vspace{1pt}

\begin{thebibliography}{9}
\bibitem {1} M. A. Olshanetsky and A. M. Perelomov, Phys. Rep. 71 (1981), 313.

\bibitem {2} J. Avan and T. Talon, Phys. Lett. B303 (1993), 33.

\bibitem {3} F. Calogero, Lett. Nuovo Cim. 13 (1975), 411; 16 (1976), 77.

\bibitem {4} J. Moser, Adv. Math. 16 (1975), 1.

\bibitem {5} I. M. Krichever, Func. Anal. Appl. 14 (1980), 282.

\bibitem {6} E. K. Sklyanin, Func. Anal. Appl. 16 (1982), 263; 17 (1983),
320; Commun. Math. Phys. 150 (1992), l81; hep-th/9308060.

\bibitem {7} H. W. Braden and T. Suzuki, Lett. Math. Phys. 30 (1994), 147.

\bibitem {8} O. Babelon and D. Bernard, Phys. Lett. B317 (1993), 363.

\bibitem {9} H. W. Braden, Andrew and N. W. Hone, Phys. Lett. B380 (1996), 296.

\bibitem {10} B. Y. Hou and W. L. Yang, Lett. Math. Phys. 44 (1998), 35; J.
of Phys. A32 (1999), 1475.

\bibitem {11} B. Y. Hou and W. L. Yang, IMPNWU-971219; math.QA/9802104.

\bibitem {12} J. Avan, O. Babelon and E. Billey Commun. Math. Phys. 178
(1996), 281.

\bibitem {13} H. W. Braden and R. Sasaki, Prog. Theor. Phys. 97 (1997), 1003.

\bibitem {14} H. W. Braden, E. Corrigan, P. E. Dorey and R. Sasaki, Nucl.
Phys. B338 (1990), 689; B356 (1991), 469.

\bibitem {15} F. W. Nijhoff O. Ragnisco and V. B. Kuznetsov, Commun. Math.
Phys. 176 (1996), 681.

\bibitem {16} G. E. Arutyunov, S. A. Frolov and P. B. Medredev, J. of Phys.
A30 (1997), 5051; J. Math. Phys. 38 (1997), 5682.

\bibitem {17} A. Gorsky and N. Nekrosov, Nucl. Phys. B414 (1994), 213; B436
(1995), 582.

\bibitem {18} S. N. M. Ruijsenaars, Commun. Math. Phys. 115 (1988), 127.

\bibitem {19} Yuri B. Suris, hep-th/9603011.

\bibitem {20} F. W. Nijhoff V. B, Kuznetsov, E. K. Sklyanin and O. Ragnisco,
J. of Phys. A29 (1996), L333.

\bibitem {21} A. J. Bordner, E. Corrigan and R. Sasaki, Prog. Theor. Phys.
100 (1998), 1107.

\bibitem {22} A. J. Bordner, R. Sasaki and K. Takasaki, Prog. Theor. Phys.
101 (1999), 487.

\bibitem {23} A. J. Bordner and R. Sasaki, Prog. Theor. Phys. 101 (1999), 799.

\bibitem {24} A. J. Bordner, E. Corrigan and R. Sasaki, Prog. Theor. Phys.
102 (1999), 499; hep-th/9905011.

\bibitem {25} O. Babelon and D. Bernard, Phys. Lett. B317 (1993), 363.

\bibitem {26} L. Freidel and J. M. Maillet, Phys. Lett. B262 (1991), 278.

\bibitem {27} S. N. M. Ruijsenaars and H. Schneider, Ann. of Phys. 170
(1986), 370.

\bibitem {28} A. A. Belavin, Nucl. Phys. B180 (1980), l09.

\bibitem {29} M. P. Richey and C. A. Tracy J. Stat. Phys. 42 (1986), 311.

\bibitem {30} M. Jimbo, T. Miwa and M. Okado, Nucl. Phys. B300 (1988), 74.

\bibitem {31} B. Y. Hou, K. J. Shi and W. L. Yang, J. of Phys. A26 (1993), 4951.

\bibitem {32} B. Y. Hou and H. Wei, J. Math. Phys. 3O (1989), 2750.

\bibitem {33} K. Hasegawa, J. of Phys. A26 (l993), 3211; J. Math. Phys. 35
(1994), 6158.

\bibitem {34} Y. H. Quano and A. Fujii, Mod. Phys. Lett. A6 (1991), 3635.

\bibitem {35} B. Y. Hou, K. J. Shi, W. L. Yang and Z. X. Yang, Phys. Lett.
A178 (1993), 73.

\bibitem {36} K. Hasegawa, q-alg/9512029; Commun. Math. Phys. 187 (1997), 289.

\bibitem {37} E. D'Hoker and D. H. Phong, hep-th/9808156.

\bibitem {38} K. Hikami and Y. Komori, J. Phys. Soc. Jpn. 67 (1998), 4037.

\bibitem {39} G. Felder and A. Varchenko, q-alg/9605024; Nucl. Phys. B480
(1996), 485.

\bibitem {40} R. H. Yue et al., Preprint (NWU-IMP9906).

\bibitem {41} E. K. Sklyanin and T. Takebe, Phys. Lett. A219 (1996), 217; solv-int/9807008.

\bibitem {42} K. Chen, B. Y. Hou, W. L. Yang and Y. Zhen, Chinese. Phys.
Lett. 16 (1999), 1.
\end{thebibliography}
\end{document}